\documentclass[aps,10pt,nofootinbib,groupedaddress,preprintnumbers,superscriptaddress]{revtex4}

\usepackage{amssymb, amsfonts, amsmath, bm}
\usepackage[
]{graphicx}
\usepackage{color}
\usepackage{ulem}

\usepackage[
]{hyperref}
\hypersetup{%
 setpagesize=false,
 bookmarksnumbered=true,%
 bookmarksopen=true,%
 colorlinks=true,%
 linkcolor=blue,%
 citecolor=red}

\begin{document}
\title{
{\large
Innermost stable circular orbits
in Majumdar--Papapetrou dihole spacetime

}}
\author{{\large Keisuke Nakashi}}
\email{nakashi@rikkyo.ac.jp}
\affiliation{Department of Physics, Rikkyo University, Toshima, Tokyo 171-8501, Japan}
\author{{\large Takahisa Igata}}
\email{igata@rikkyo.ac.jp}
\affiliation{Department of Physics, Rikkyo University, Toshima, Tokyo 171-8501, Japan}
\date{\today}
\preprint{RUP-19-10}

\begin{abstract}
We investigate the positions of stable circular massive particle orbits 
in the Majumdar--Papapetrou dihole spacetime with equal mass.
In terms of qualitative differences of 
their sequences,
we classify the dihole separation into five ranges and 
find four critical values as the boundaries.
When the separation is relatively large, 
the sequence on the symmetric plane bifurcates, 
and furthermore, they extend to each innermost stable circular orbit 
in the vicinity of each black hole.
In a certain separation range, 
the sequence on the symmetric plane separates into two parts.
On the basis of this phenomenon, 
we discuss the formation of double accretion disks with a common center.
Finally, we clarify the dependence of the radii
of marginally stable circular orbits and innermost stable circular orbits 
on the separation parameter.
We find a discontinuous transition of the innermost stable circular orbit radius. 
We also find 
the separation range at which 
the radius of the innermost stable circular orbit can be smaller than that of the stable circular photon orbit. 
\end{abstract}
\maketitle 

\section{Introduction}
\label{I}

In 2015, the LIGO Scientific and Virgo Collaborations 
detected gravitational waves from 
a binary black hole~\cite{Abbott:2016blz,TheLIGOScientific:2016wfe}. 
This observation established that the binary black hole systems exist in nature.
So far, we have already found ten binary black hole mergers 
and one binary neutron star merger~\cite{GBM:2017lvd, LIGOScientific:2018mvr}, 
and the number of detections will increase further in the future. 
These discoveries strongly motivate the 
study of phenomena around a binary black hole system.

Actual binary black hole systems are so dynamic that
the method of numerical relativity is quite useful to 
describe phenomena in these systems. 
On the other hand, it is also significant
to use analytical methods 
for a qualitative
understanding of the phenomena.
However, there is no analytical expression 
of a binary black hole system due to their dynamical character. 
Therefore, we often adopt a static (or stationary) and axisymmetric 
dihole spacetime as a toy model.
There are some dihole solutions 
of the Einstein equation (or the Einstein--Maxwell equation) 
with these symmetries, 
such as the Weyl spacetime~\cite{Weyl:1917gp}, 
the Majumdar--Papapetrou 
spacetime~\cite{Majumdar:1947eu, Papaetrou:1947ib, Hartle:1972ya}, 
the double-Kerr spacetime~\cite{Kramer:1980}, etc. 
We can learn a lot from phenomena on such a
spacetime and obtain strong suggestions to binary black hole events.
For example, while the formation of binary black hole shadows 
requires fully nonlinear simulation of numerical relativity, 
we can identify some specific features 
by using a (quasi-)static dihole 
spacetime~\cite{Nitta:2011in, Bohn:2014xxa, Patil:2016oav,
Assumpcao:2018bka,
Cunha:2018cof}.

The research of test particle motion in 
strong gravitational fields
is significant in astrophysics as well as in gravitational theories.
Even in binary black hole systems, 
it is still one of the most fundamental problems. 
Indeed, the formation of the 
binary black hole shadow mentioned above is 
the problem of the dynamics of massless particles.
On the other hand, the dynamics of massive particles in a binary black hole system 
has been discussed in the contexts of gravitational wave radiation 
induced by a third body 
effect~\cite{Campanelli:2005kr, Torigoe:2009bw, Seto:2012ig, Yamada:2015abm} 
and the formation of 
multiple accretion disks~\cite{Hayasaki:2006fq,Kimitake:2007fs}. 
In these phenomena, 
the sequence of stable circular orbits is crucial, 
and in particular, 
the innermost stable circular orbit~(ISCO) plays a key role 
because 
it is expected to be the inner edge of an accretion disk~\cite{Novikov:1973},
and also an inspiralling compact binary transits into 
the merging phase there~\cite{Clark:1977,Kidder:1993zz}.

Under these circumstances, 
the purpose of this paper is to clarify the characteristics 
of stable circular orbits in a binary black hole system.
In particular, it is meaningful to study 
how marginally stable circular orbits (or innermost stable circular orbits) 
of a binary black hole system changes compared to those of a single black hole.
To achieve this, we consider stable circular orbits 
around the axis of symmetry of the dihole
in the equal mass Majumdar--Papapetrou~(MP) spacetime,
which is an exact solution of the Einstein--Maxwell equation 
and consists of two extremal Reissner--Nordstr\"om black holes. 
Once we fix the mass scale of the dihole, 
it depends only on one parameter, the dihole separation.
We expect that the stable circular orbits exhibit 
nontrivial appearance when the separation between two black holes changes.
Indeed, in some previous works, 
several properties of stable circular orbits in the MP dihole spacetime were reported: 
marginally stable circular orbits of massive particles 
are not unique~\cite{Wunsch:2013st}
and the stable circular photon orbit appears 
inside the unstable circular photon orbit~\cite{Dolan:2016bxj}.  
In this paper, 
we analyze the dependence of the sequence of stable circular orbits 
on the separation parameter in the whole spacetime. 
As a result of our analysis, 
the parameter range is divided into five, 
at each boundary of which the behavior of stable circular orbits drastically changes. 
Then we find the appearance of 
marginally stable circular orbits, ISCOs, and the stable/unstable circular photon orbit. 
Note that the marginally stable circular orbit 
is unique in the Schwarzschild/Reissner--Nordstr\"om/(charged) Kerr spacetime
but not in the MP dihole spacetime. 
Therefore, the ISCO is the marginally stable circular orbit with the smallest radius.
Finally, we find that the position of the ISCO transits discontinuously 
at a certain value of the separation 
and the radius of the ISCO is smaller than the 
stable circular photon orbit
for a specific separation.

This paper is organized as follows. 
In Sec.~\ref{II}, 
we derive the conditions of stability for 
circular orbits of a massive particle 
in the MP dihole spacetime with equal mass. 
These conditions are given in terms of 
a 2D effective potential and its Hessian. 
In Sec.~\ref{III}, 
we explore 
the dependence of the sequence of stable circular orbits 
on the separation between the dihole. 
Dividing the range of the separation parameter into 
five parts, 
we clarify the transitions of marginally stable circular orbits 
and the innermost stable circular orbit. 
Furthermore, by analyzing 
qualitative changes of the sequence of 
stable circular orbits, 
we find some critical values of the separation parameter analytically.
Section~\ref{IV} is devoted to a summary and discussions. 
We use units in which 
$G=1$ and $c=1$.

\section{Stability conditions of 
circular orbits in the Majumdar--Papapetrou dihole spacetime with equal mass}
\label{II}
The metric and the gauge field of the Majumdar--Papapetrou~(MP) 
dihole spacetime in isotropic coordinates are given by 
  \begin{align}
   g_{\mu \nu } \mathrm{d} x^\mu \mathrm{d}x^\nu
   &= - \frac {\mathrm{d}t ^ 2} {U ^2} 
   + U ^ { 2 } \left(\mathrm{d}  \rho ^ { 2 } 
   + \rho ^ { 2 } \mathrm{d} \phi ^ { 2 } 
   + \mathrm{d} z ^ { 2 } \right), 
   \label{metric}\\
    A_\mu \mathrm{d}x^\mu &= U^{-1} \mathrm{d}t, \\
    U(\rho, z ) 
    &= 1+ \frac{M_+}{\sqrt{\rho ^2 + (z- a)^2}}
    +\frac{M_-}{\sqrt{\rho ^2 + (z+ a)^2}},
  \end{align}
where $M_\pm$ are masses of two extremal Reissner--Nordstr\"om black holes 
located at $z=\pm a\  (a\geq 0)$. 
Note that we choose cylindrical coordinates on the spatial geometry, 
$x=\rho \cos \phi$ and $y=\rho \sin \phi$, 
where $x$, $y$ are the Cartesian coordinates.

The Lagrangian $\mathcal{L}$ of a freely falling
test particle is defined by 
  \begin{align}
    \mathcal{L} = \frac{1}{2}g_{\mu \nu } \dot x^\mu \dot x^\nu 
    = \frac{1}{2} \left [ -\frac{\dot t^2}{ U^2}  
    + U^2 (\dot \rho ^2 +\rho ^2 \dot \phi ^2 + \dot z^2) \right ],
  \end{align}
where the dot denotes derivative with respect to an affine parameter. 
Since the spacetime is static and axisymmetric, 
the Lagrangian does not depend on the coordinates $t$ and $\phi$ explicitly.
Thus, there are two constants of motion
  \begin{align}
    E = \frac{\dot t}{U^2},\ \ \ 
    L=\rho ^2 U^2\dot \phi,
    \label{EL}
  \end{align}
where we may assume that $E>0$. These quantities
correspond to 
energy and angular momentum,
respectively. 
Without loss of generality, we assume that the worldline is parametrized 
so that $g_{\mu \nu}\dot x^\mu \dot x^\nu = -\kappa$, where $\kappa =1$ for timelike particles and $\kappa =0$ for null particles. 
From this normalization and Eqs.~\eqref{EL}, 
we introduce
the effective potential $V(\rho, z) $ for the 
motion in $\rho$-$z$ plane 
  \begin{align}
  \label{eq:normalization}
    &\dot \rho ^2 + \dot z ^2+V= E^2, \\
    &V(\rho, z; L^2) = \frac{L^2 }{\rho ^2 U^4} + \frac{\kappa }{U^2}.
    \label{potential}
  \end{align}

We consider circular orbits with constant $\rho$ and $z$ 
for timelike particles.
Particles in circular orbits must satisfy the following 
conditions: (a) $\dot \rho = \dot z =0$, (b) $\ddot \rho = \ddot z = 0$.
Condition~(a) together with the normalization~\eqref{eq:normalization} leads to
\begin{align}
\label{eq:V=E^2}
V = E^2.
\end{align}
Conditions~\eqref{eq:V=E^2}, (a), and (b) together 
with the equations of motion for $z$ and $\rho$ imply
  \begin{align}
    V_{z}  = 0, \label{Vz} \\
    V_{\rho } = 0, \label{Vr}
  \end{align}
where $V_i = \partial _i V \ (i=z, \rho)$.
Hence the circular orbits are realized at stationary points of $V$ 
where the values of $V$ are positive.

We can rewrite Eqs.~\eqref{eq:V=E^2}, \eqref{Vz}, and \eqref{Vr} 
by using the explicit form~\eqref{potential} as 
\begin{align}
\label{Uz}
&U_z=0,
\\
\label{L0}
&L^2=L_0 ^2(\rho, z)
:= - \frac{\rho ^3 U^2 U_\rho}{U + 2\rho\:\! U_\rho},
\\
\label{E0}
&E^2=E_0^2(\rho, z):=V(\rho, z; L_0^2),
\end{align}
where $L_0 ^2$ must be 
non-negative value in the physical branch. 
Note that, if $L_0^2\geq 0$ holds, 
then $V$ is necessarily positive. 
Finally, we find that a particle moves in a circular orbit with constant $\rho$ and $z$ 
if and only if the conditions~$U_z=0$, $L^2=L_0^2\geq0$, and 
$E^2=E_0^2$
are satisfied.

We also consider 
the stability of circular orbits. 
From the standard linear stability analysis of circular orbits, 
we find that 
a circular orbit is stable if and only if 
the orbit exists at a local minimum point of $V$. 
We call such a circular orbit 
a stable circular orbit. 
On the other hand, 
a circular orbit is unstable if and only if 
the orbit exists at a local maximum point of $V$ or 
a saddle point of $V$. 
We call such a circular orbit 
an unstable circular orbit. 
Here we 
introduce the Hessian of $V$ and the trace of $V_{ij}=\partial_i \partial_j V$
\begin{align}
&h(\rho, z; L^2) = \det V_{ij},
\\
&k(\rho, z; L^2)=\mathrm{Tr}\:\! V_{ij}.
\end{align}
In terms of these quantities, we can summarize 
the stability of circular motions as follows:
\begin{itemize}
\item[(i)] A circular orbit is stable 
$\Longleftrightarrow$ 
$h>0$ and $k>0$ at a stationary point of $V$;
\item[(ii)] A circular orbit is unstable
$\Longleftrightarrow$ 
($h >0$ and $k < 0$) or $h<0$ at a stationary point of $V$.
\end{itemize}
When a sequence of stable circular orbits switches to a sequence of unstable circular orbits 
at a radius, 
we call the circular orbit at the radius a marginally stable circular orbit, 
where $V$ has an inflection point (i.e., $h=0$).
In particular, we
call the marginally stable circular orbit with 
the smallest value of the radial coordinate $\rho$ 
the innermost stable circular orbit.

In the remainder of this paper, we investigate circular orbits 
in the MP dihole spacetime 
with equal mass 
$M_+=M_-$. We use units in which $M_\pm=1$ in what follows.
In this case, Eq.~\eqref{Uz} reduces to
  \begin{align}
    z \left[ \rho ^ { 6 } 
    - 3 \left( a ^ { 2 } - z ^ { 2 } \right) ^ { 2 } \rho ^ { 2 } 
    - 2 \left( a ^ { 2 } + z ^ { 2 } \right) 
    \left( a ^ { 2 } - z ^ { 2 } \right) ^ { 2 } \right] = 0.
    \label{modiUz}
  \end{align}
This equation means that
$U_z$ always vanishes on the symmetric plane $z=0$.
Focusing on the case where the inside of the square bracket vanishes, 
we find another real root of Eq.~\eqref{Uz}
  \begin{align}
    \rho_0 ^2 = 2(a^2 - z^2)\cos 
    \left [ \frac{1}{3} \arccos \frac{a^2 + z^2 }{a^2 - z^2} \right ].
    \label{slncurve}
  \end{align}
Hence we obtain two curves $z=0$ and $\rho=\rho_0(z)$ in $\rho$-$z$ plane, where 
$U_z$ vanishes.
Note that these curves intersect each other at $(\rho, z)=(\sqrt{2}a, 0)$.
To discuss the stability of circular orbits,
we introduce the Hessian $h$ and the trace $k$ evaluated at $L^2 = L_0 ^2$
\begin{align}
    h_0(\rho, z) = h(\rho, z; L_0^2)|_{U_z=0},
    \\
k_0(\rho, z)=k(\rho, z; L_0^2)|_{U_z=0},
\end{align}
where the restriction $U_z=0$ means that 
the terms directly proportional to $U_z$ have been removed from the right-hand sides.
Using $h_0$ and $k_0$, we specify the region where the remaining conditions for 
stable circular orbits hold
\begin{align}
D=\{(\rho, z) \,|\, h_0>0, k_0>0, L_0^2>0\}.
\end{align}
Hence, stable circular orbits in the MP dihole spacetime with equal mass 
exist on the curves $z=0$ or $ \rho=\rho_0$ included in the region $D$ in $\rho$-$z$ plane.

It is convenient to obtain the expressions 
$L_0$, $E_0$, and $h_0$ evaluated at 
the symmetric plane $z=0$. To derive them
in simpler forms, 
we use a new coordinate $R$ defined by
\begin{align}
R(\rho)=\sqrt{\rho^2+a^2},   
\end{align}
where $R\geq a$, which follows from $\rho \geq 0 $. 
In terms of $R$, we 
derive angular momentum and energy for a circular orbit on $z=0$, respectively,
\begin{align}
  \label{eq:L0z=0}
L _ { 0 }(\rho,
   0 ) 
&= \frac{\sqrt{2}\:\! ( R + 2 )( R^2 - a ^2)}{R\sqrt{f}},
\\
E_0(\rho,0)
&=\frac{R\sqrt{R^3+2a^2}}{(R+2)\sqrt{f}},
\end{align}
where, 
without loss of generality, we have chosen the branch 
$L_0\geq0$, and  
\begin{align}
\label{eq:f}
f(R)&=R^3 -2 R^2 + 4a^2.
\end{align}
Note that these quantities diverge 
if $f$ vanishes. 
Furthermore, 
in the range $f<0$, there is no circular orbit~(see Sec.~\ref{sec:III_F} for details).
In addition, 
the derivatives of $L_0(\rho, 0)$ and $E_0(\rho, 0)$ with respect to $R$
are given by, respectively,
\begin{align}
\label{eq:dL0/dR}
\frac{\mathrm{d}L_0(\rho, 0)}{\mathrm{d}R}
&=
\frac{g}{\sqrt{2}R^2 f^{3/2}},
\\
\label{eq:dE0/dR}
\frac{\mathrm{d}E_0(\rho, 0)}{\mathrm{d}R}
&=\frac{g}{(R+2)^2\sqrt{R^3+2a^2}f^{3/2}},
\end{align}
where 
\begin{align}
\label{eq:g}
g(R)&=R^6-6 R^{5}+3a^2R^4+22a^2 R^{3}+16a^4.
\end{align}
These results mean that 
the monotonicity of angular momentum and energy for a circular orbit switches at 
the points where $g=0$. 
We find that, 
at least in the region far enough from the center $R\gg a$, 
the angular momentum $L_0(\rho, 0)$ 
and the energy $E_0(\rho,0)$ are real positive values and 
monotonically increasing with $R$. 
We also derive $h_0$ evaluated at $z=0$ 
\begin{align}
h_0(\rho, 0)=
\frac{16\:\! (R^2-3\:\!a^2)\:\!g}{R^2\:\!(R+2)^6\:\!f^2}.
\label{h0onz=0}
\end{align}
If $f$ vanishes, 
this quantity diverges, 
which is similar to the behaviors seen 
in $L_0(\rho,0)$ and $E_0(\rho,0)$. 
On the other hand, 
the Hessian $h_0(\rho,0)$ vanishes at $R=\sqrt{3}a$ (i.e., $\rho=\sqrt{2}a$), where 
$z=0$ and $ \rho=\rho_0$ intersect each other. 
In addition, 
$h_0(\rho, 0)$ also vanishes for $g=0$, which is 
similar to the 
behavior seen in Eqs.~\eqref{eq:dL0/dR} and \eqref{eq:dE0/dR}. 
This fact implies that 
the monotonicity of angular momentum and energy for a circular orbit switches at 
zeros of $h_0(\rho, 0)$. 
Taking into account $h_0(\rho, 0)>0$ for $R\gg a$, 
we find that angular momentum and energy 
monotonically increase with $R$  
on the sequence of 
stable circular orbits 
and monotonically decrease with $R$
on the sequence of 
unstable circular orbits.

\begin{figure}[t]
\centering
\includegraphics[clip,width=18.0cm]{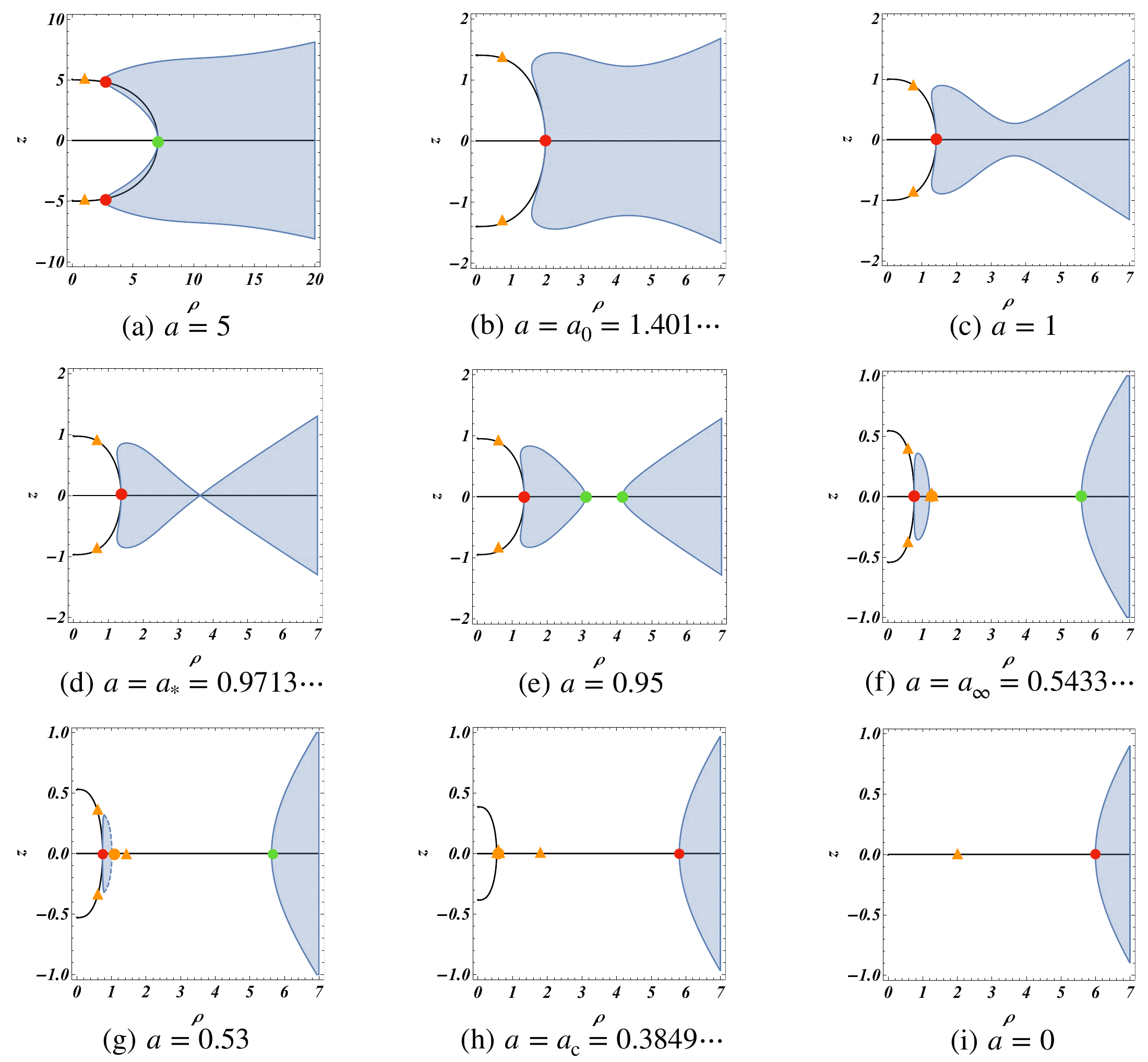}
    \caption{Positions of stable circular orbits 
    of massive particles 
    in $\rho$-$z$ plane of
    the Majumdar--Papapetrou dihole spacetime 
    with equal mass $M_+=M_-=1$ 
    for the separation range $0 \leq a \leq 5$. 
    The solid black lines represent 
    the curves satisfying $U_z=0$ (i.e., $z=0$ and $\rho=\rho_0$). 
    The shaded regions show the region $D$, 
    where $h_0>0$, $k_0>0$, and $L_0^2 >0$ are satisfied. 
    The sequence of stable circular orbits is the solid black curves included in the region $D$. 
    The solid blue lines are 
    the boundary of $D$ where $h_0=0$, $L_0^2>0$, and $k_0>0$.
    The dashed blue lines are 
    the boundary of $D$ where $h_0>0$, $k_0>0$, and $L_0^2$ diverges.
    The red dots are the positions of ISCOs.
    The green dots are the positions of 
    marginally stable circular orbits~(MSCOs) 
    except for the ISCO. 
    The orange triangles and dots 
    are the positions of unstable circular photon orbits and stable 
    ones, where
    $L_0^2$ diverges. 
    (a) $a=5$: When $a$ is large enough, 
    stable circular orbits exist not only on $z=0$ line in the range $\rho\in (\sqrt{2}a, \infty)$ 
    but on $\rho=\rho_0$ line. An MSCO exists at $(\rho, z)=(\sqrt{2}a,0)$, and 
    the ISCOs are located around each black hole. 
    (b) $a=a_0=1.401\cdots$: There exist stable circular orbits on $z=0$ plane 
    in the range $\rho\in (\sqrt{2}a_0, \infty)$. The ISCO is located at $(\rho, z)=(\sqrt{2}a_0, 0)$, 
    where three MSCOs for $a>a_0$ are degenerate.
    (c) $a=1$: There exist stable circular orbits on $z=0$ plane in the range $\rho\in (\sqrt{2},
    \infty)$. The point $(\rho, z)=(\sqrt{2},
    0)$ is the ISCO.
    (d) $a=a_*=0.9713\cdots$: The region $D$ is marginally connected at $(\rho, z)=(\rho_*, 0)$. 
    There exist stable circular orbits on $z=0$ plane 
    in the range $\rho\in (\sqrt{2}a_*, \infty)$, of which the boundary $(\rho, z)=(\sqrt{2}a_*, 0)$ is the ISCO.
    (e) $a=0.95$: 
    The sequence of stable circular orbits 
    splits into two parts. 
    As a result, 
    two additional MSCOs appear at the 
    boundary of the outer sequence and the outer boundary of the inner sequence.
    The point $(\rho, z)=(\sqrt{2}a, 0)$ is the ISCO.
    (f) $a=a_\infty=0.5433\cdots$:
    There are two sequences of stable circular orbits. 
    The outer boundary of the inner sequence is 
    no longer physical because $L_0^2$
    diverges there. An MSCO appears at the boundary of the outer sequence, 
    and the ISCO appears at $(\rho, z)=(\sqrt{2}a_\infty, 0)$. 
    (g) $a=0.53$: There are two sequences of stable circular orbits. 
    At the outer boundary of the inner sequence, 
    $L_0^2$ diverges.
    An MSCO appears at the boundary of the outer sequence, 
    and the ISCO appears at $(\rho, z)=(\sqrt{2}a, 0)$.
    (h) $a=a_{\textrm{c}}=0.3849\cdots$:
    The divergence of $L_0^2$ occurs 
    at $(\rho, z)= (\sqrt{2}a_{\textrm{c}}, 0)$
    and the inner sequence of stable circular orbits disappears. 
    There only exists the outer sequence and its inner boundary becomes the ISCO. 
    (i) $a=0$: There exist stable circular orbits in the range $\rho\in(6, \infty)$ on $z=0$.
    The point $(\rho, z)=(6, 0)$ is the ISCO, which 
    connects to the ISCO of the single extremal Reissner--Nordstr\"om black hole with mass 2. } 
    \label{positions} 
\end{figure}

\section{Dependence of the sequence of stable circular orbits on the separation}
\label{III}
We discuss the dependence of the positions of stable circular orbits on 
the separation parameter $a$. 
Using the functions defined in the previous section, 
we plot the sequence of stable circular orbits as illustrated in Figs.~\ref{positions}. 
On the basis of these plots, 
dividing the range of $a$ into five parts, 
we clarify the behavior of stable circular orbits for each range of $a$ in the following subsections. 
Furthermore, we find the four critical values of $a$ 
characterized by the behaviors of the sequence of stable circular orbits 
and the angular momentum of a circular orbit.

\subsection{$a>1.401\cdots$}
We focus on stable circular orbits in the case where
the separation between the dihole is large enough (i.e., $a\gg1$).
Figure~\ref{positions}(a) shows 
a typical shape of the sequence of stable circular orbits for a large value of $a$.
As seen from the figure, 
stable circular orbits exist on the line $z=0$ 
in the range $\rho\in (\sqrt{2}a, \infty)$. 
The end point $(\rho, z)=(\sqrt{2}a,0)$ is a marginally stable circular orbit 
because the sequence switches to that of unstable circular orbits at this point,
where $h_0=0$.
In addition, at this point the sequence of stable circular orbits bifurcates into 
the curve $\rho=\rho_0$, where $\rho_0$ is defined by Eq.~\eqref{slncurve}. 
Finally it terminates near each black hole, 
which also correspond to 
marginally stable circular orbits, especially the ISCOs. 

Even in the sequence on $\rho=\rho_0$,
the energy and the angular momentum of stable circular orbits 
monotonically decrease as the radius decreases up to the ISCOs.
Note that, when $a$ is large enough, 
a particle moving near each black holes feels
gravity of a single black hole. 
Indeed, 
in the limit as $a\to \infty$,
the ISCO radius measured by $\rho$ approaches $3$, 
which
coincides with the ISCO radius of the single extremal Reissner--Nordstr\"om 
black hole spacetime~(see Appendix~\ref{sec:B}).

As the value of $a$ decreases from a large value, 
the ISCOs approach the intersection of $z=0$ line and $\rho=\rho_0$ line. 
When the value of $a$ reaches $1.401\cdots$, 
the three marginally stable circular orbits merge at 
a point on $z=0$~[see Fig.~\ref{positions}(b)]. 
As a result, the sequence of stable circular orbits only appears on the line $z=0$.

\subsection{$a=a_0=1.401\cdots$}

We find the critical value 
$a=a_0$
at which 
the three marginally stable circular orbits degenerate~[see Fig.~\ref{positions}(b)]. 
We expand $\rho_0$ in Eq.~\eqref{slncurve} around $z=0$ up to $O(z^2)$, 
  \begin{align}
    \rho _0 = \sqrt{2}a - \frac{7}{9\sqrt{2}a}z^2 + O(z^4).
  \end{align}
Substituting this expression into $h_0$, we expand it around $z=0$ again,
  \begin{align}
    h_0 (\rho _0, z) 
    = \frac{768 \left(54 a^2-33 \sqrt{3} a-26\right) }{a^2\left(9 a-2\sqrt{3}\right)^2 \left(3 a+2\sqrt{3}\right)^6}z^2+O(z^4).
  \end{align}
As already discussed above, these results imply that 
there exists a marginally stable circular orbit at the point $(\rho, z)=(\sqrt{2}a, 0)$. 
Furthermore, since the condition of the multiple root is 
$\mathrm{d}^2h_{0}(\rho_0, z)/\mathrm{d}z^2 = O(z^2)$,
i.e., $54 a^2-33 \sqrt{3} a-26=0$, 
we obtain the critical value $a_0$ as
\begin{align}
    a_0 = \frac{11+\sqrt{329}}{12\sqrt{3}} = 1.401\cdots.
    \label{a0}
\end{align}
Thus, in the case $a=a_0$, 
we find stable circular orbits on $z=0$ plane in the range
$\rho\in (\sqrt{2}a_0, \infty)$ and the ISCO at $(\rho, z)=(\sqrt{2}a_0, 0)$.

\subsection{$a_0>a>0.9713\cdots$}
If we make $a$ smaller than $a_0$, 
the sequence of 
stable circular orbits still appears 
only on $z=0$ plane in the range $\rho\in (\sqrt{2}a, \infty)$~[see Fig.~\ref{positions}(c)],
so that it is sufficient to analyze circular orbits on it.
The end point $(\rho, z)=(\sqrt{2}a,0)$
corresponds to the unique 
marginally stable circular orbit, especially the ISCO.
When the value of $a$ reaches $0.9713\cdots$, 
the region $D$ becomes marginally connected at the intersection point 
of the lines $z=0$ and $h_0=0$~[see Fig.~\ref{positions}(d)].  
This intersection point is not a 
marginally stable circular orbit 
because the sequence of stable circular orbits does not switch to 
that of unstable circular orbits here.
As a result, there exists the unique marginally stable circular orbit 
in the range $a_0\geq a \geq 0.9713\cdots$.

\subsection{$a=a_*=0.9713\cdots$}
We seek the exact critical value $a=a_*$ at which 
the region $D$ is 
marginally connected at the intersection point of
the lines $z=0$ and $h_0=0$~[see Fig.~\ref{positions}(d)].
In other words, 
the function $h_0$ has a saddle point at 
this point. 
We use this condition to derive $a_*$ in what follows. 
From the explicit form of $h_0(\rho, 0)$ given in Eq.~\eqref{h0onz=0}, 
we find that the condition $h_0(\rho, 0)=0$ holds at $(R, z)=(\sqrt{3}a, 0)$,
but
the Hessian $h_0$ does not have a stationary point there. 
Therefore we focus on the other branch
\begin{align}
\label{eq:R6}
g=0,
\end{align}
where $g$ is defined by Eq.~\eqref{eq:g}.
A point satisfying this equation can be 
a stationary point of $h_0$ if 
$\mathrm{d}h_{0}(\rho,0)/\mathrm{d}R=0$,
which reduces to 
\begin{align}
\label{eq:R3}
R^3-5R^2+2a^2R+11a^2=0,
\end{align}
where we have used Eq.~\eqref{eq:R6}. 
Solving Eqs.~\eqref{eq:R6} and \eqref{eq:R3} 
for $a$ and $R$ simultaneously, then we obtain 
the solutions
\begin{align}
&a_*=\frac{50 \:\!(7+\sqrt{129})}{(13+\sqrt{129})\sqrt{710+70 \sqrt{129}}}=0.9713\cdots, \label{a*}
\\
&R_*=\frac{-19+3 \sqrt{129}}{4}=3.768\cdots.
\end{align}
The value of $\rho$ corresponding to these solutions is given by
\begin{align}
\rho_*=\frac{5}{512}(20291-1667\sqrt{129})=3.641\cdots.
\end{align}
The inverse of $a_*$ 
coincides with the critical values $M_*$ mentioned in Ref.~\cite{Wunsch:2013st}.
Note that the linear stability of 
a circularly orbiting particle at $(\rho, z)=(\rho_*, 0)$ is undetermined, 
but the analysis of the allowed region of the particle motion
shows 
it nonlinearly stable.

\subsection{$a_*>a>0.5433\cdots$}
If we make $a$ smaller than $a_*$, 
the region $D$ 
is separated into two regions~[see Fig.~\ref{positions}(e)], 
and then 
two sequences of stable circular orbits appear on the line $z=0$. 
The outer sequence
appears from infinity to 
a marginally stable circular orbit, 
while the inner
sequence appears between another marginally stable circular orbit and the ISCO. 
Therefore, 
three marginally stable circular orbits appear 
in total as the boundaries of these sequences.
Their radii 
except for the ISCO radius are given as real roots for Eq.~\eqref{eq:R6}, 
and the ISCO radius is $\rho = \sqrt{2}a$ in particular.

As the value of $a$ gradually decreases, 
the two sequences tend to separate from each other. 
In addition, the energy at the marginally stable circular orbit 
next to the ISCO increases. 
Remarkably, it reaches the energy level of a rest particle at infinity (i.e., $E=1$)
at $a=0.7567\cdots$.
Hence, for $a\leq 0.7567\cdots$,
stable circular orbits with $E_0\geq1$ exist
until the inner sequence disappears.
Note that we do not observe such a phenomenon 
in the Kerr spacetime. 
Since circular orbits with $E_0\leq 1$
occur more naturally,
the sequence with $E_0>1$ 
does not contribute to 
phenomena such as accretion disk formation around the dihole.

When the value of $a$ reaches $0.5433\cdots$, 
the marginally stable circular orbit next to the ISCO
is no longer a circular orbit because 
infinitely large angular momentum and energy are 
required to keep it a circular orbit~[see Fig.~\ref{positions}(f)]. 
In the following subsection, 
we find the critical value of $a$ 
from the behavior of $L_0^2$.

\subsection{$a=a_\infty=0.5433\cdots$}
\label{sec:III_F}
As mentioned in the previous subsection, 
one of the three
marginally stable circular orbits located 
next to the ISCO
disappears in the limit as $a\searrow
0.5433\cdots$.
If a timelike particle circularly 
orbited at this limiting radius for $a=0.5433\cdots$, 
the angular momentum 
$L_0^2$ would diverge.
Therefore, 
to find the exact critical value $a_\infty$ of $a$,
we analyze the behavior of 
$L_0^2(\rho, 0)$, which is given by Eq.~\eqref{eq:L0z=0}. 
Notice that this expression and Eq.~\eqref{h0onz=0} diverge if 
the following condition is satisfied: 
  \begin{align}
  \label{eq:f=0}
    f=0.
  \end{align}
It is worth pointing out that 
this condition is equivalent to 
that of the existence of circular photon orbits~(see Appendix~\ref{sec:B}). 
Since $f(R)$ has a local minimum at $R=4/3$ 
and its extreme value takes the form $f(4/3)=4(a^2 -8/27)$, 
we find that the divergence of $L_0^2$ appears only at $\rho=\rho_\infty$ for $a=a_\infty$, where 
  \begin{align}
    \label{ainf}
    &a_\infty = \frac{2\sqrt{6}}{9} = 0.5433 \cdots,
    \\
\label{eq:rhoinf}
    &\rho_\infty=\frac{2\sqrt{30}}{9}=1.217\cdots,
\end{align}
where $\rho_\infty$ corresponds to  $R=4/3$. 
At $a=a_\infty$, hence the 
inner sequence of stable circular orbits on $z=0$ plane 
exist in the range 
$\rho\in (\sqrt{2}a_\infty, \rho_\infty)$.
The inverse of 
$a_\infty$ coincides with 
$\bar{M}$
mentioned in Ref.~\cite{Wunsch:2013st}.

For $a>a_\infty$, 
the angular momentum 
$L_0^2(\rho,0)$
is positive and finite everywhere. 
This means that 
there exist stable/unstable circular orbits with arbitrary radii
on $z=0$ plane. 
On the other hand, for 
$a\leq a_\infty$, 
there exists no circular orbit of a massive particle on $z=0$ plane
in the range 
$\rho_{\mathrm{ps}}\leq\rho\leq \rho_{\mathrm{pu}}$
because $L_0^2$ can be negative or infinitely large there, where 
$\rho_{\mathrm{ps}}$ and $\rho_{\mathrm{pu}}$ 
are defined in Eqs.~\eqref{eq:rhopu} and \eqref{eq:rhops}.

\subsection{$a_\infty>a>0.3849\cdots$}
If we make $a$ smaller than $a_\infty$, 
there still exist the two sequences of stable circular orbits on $z=0$ plane~[see Fig.~\ref{positions}(g)]. 
The outer sequence 
exists from infinity to a marginally stable circular orbit. 
The inner sequence 
exists in the range 
$\rho \in (\sqrt{2}a, \rho_{\mathrm{ps}})$, 
where 
$\rho=\sqrt{2}a$
is the ISCO radius and
$\rho=\rho_{\mathrm{ps}}$
is 
the radius of the stable circular photon orbit, 
defined by Eq.~\eqref{eq:rhops} in Appendix~\ref{sec:B}.
The ISCO radius is smaller than the radius of 
the stable circular photon orbit.
Note that $L_0^2$ diverges in the limit as 
$\rho\to \rho_{\mathrm{ps}}$ 
on the inner sequence, 
which is consistent with the appearance of 
the stable circular photon orbit.

As the value of $a$ approaches $0.3849\cdots$,
the value 
$\rho_{\mathrm{ps}}$ 
approaches $\sqrt{2}a$. 
When $a=0.3849\cdots$, 
the inner sequence
disappears~[see Fig.~\ref{positions}(h)].

\subsection{$a=a_{\mathrm{c}}=0.3849\cdots$}
We seek the exact critical value of 
$a=0.3849\cdots$ at which 
the inner sequence of stable circular orbits just disappears. 
The value of $L_0^2$ at $(\rho, z)=(\sqrt{2}a,0)$ is given by
  
\begin{align}
\label{eq:L0sqatbif}
    L _ { 0 } ^ { 2 } ( \sqrt { 2 } a , 0 ) = \frac { 8 \left( a + 3 a _ { \mathrm{c} } \right) ^ { 2 } } { 3 \sqrt { 3 } \left( a - a _ { \mathrm{c} } \right) },
  \end{align}
where 
\begin{align}
a _ {\mathrm{c}}
=\frac{2\sqrt{3}}{9}
= 0.3849 \cdots. 
\label{ac}
\end{align}
This result together with Eq.~\eqref{h0onz=0} means that, 
even if $a$ arbitrarily approaches to $a_{\textrm{c}}$ from above, 
the point $(\rho, z)=(\sqrt{2}a,0)$
is necessarily a marginally stable circular orbit. 
If $a=a_{\mathrm{c}}$, 
then $L_0^2$ at $(\rho, z)=(\sqrt{2}a_{\mathrm{c}},0)$ diverges, so that the inner sequence of stable circular orbits disappears. 
Consequently, we can identify $a_{\mathrm{c}}$ with the numerical critical value 
$a=0.3849 \cdots$.
Thus the location of the ISCO changes discontinuously at $a=a_{\mathrm{c}}$.
Note that, however, the 
circular photon orbit exists there.

\subsection{$a_{\mathrm{c}}>a\geq0$}
If we make $a$ smaller than $a_{\textrm{c}}$, 
the single sequence of stable circular orbits appears 
on $z=0$ plane
from infinity to the ISCO.
As the value of $a$ approaches 0, 
the ISCO radius monotonically increases. 
For $a=0$, 
the MP dihole becomes the single 
extremal Reissner--Nordstr\"om black hole 
with mass equal to 2 in our units.
Then the sequence of stable circular orbits exists from infinity to 
the ISCO radius equal to three times its mass~(see Appendix~\ref{sec:A}). 
Therefore, we find 
the ISCO at $\rho=6$ as shown in Fig.~\ref{positions}(i).
Note that $z=0$ plane at $a=0$ is 
no longer special because spherical symmetry is restored.

\begin{figure}[t]
\centering
    \includegraphics[clip,width=15.0cm]{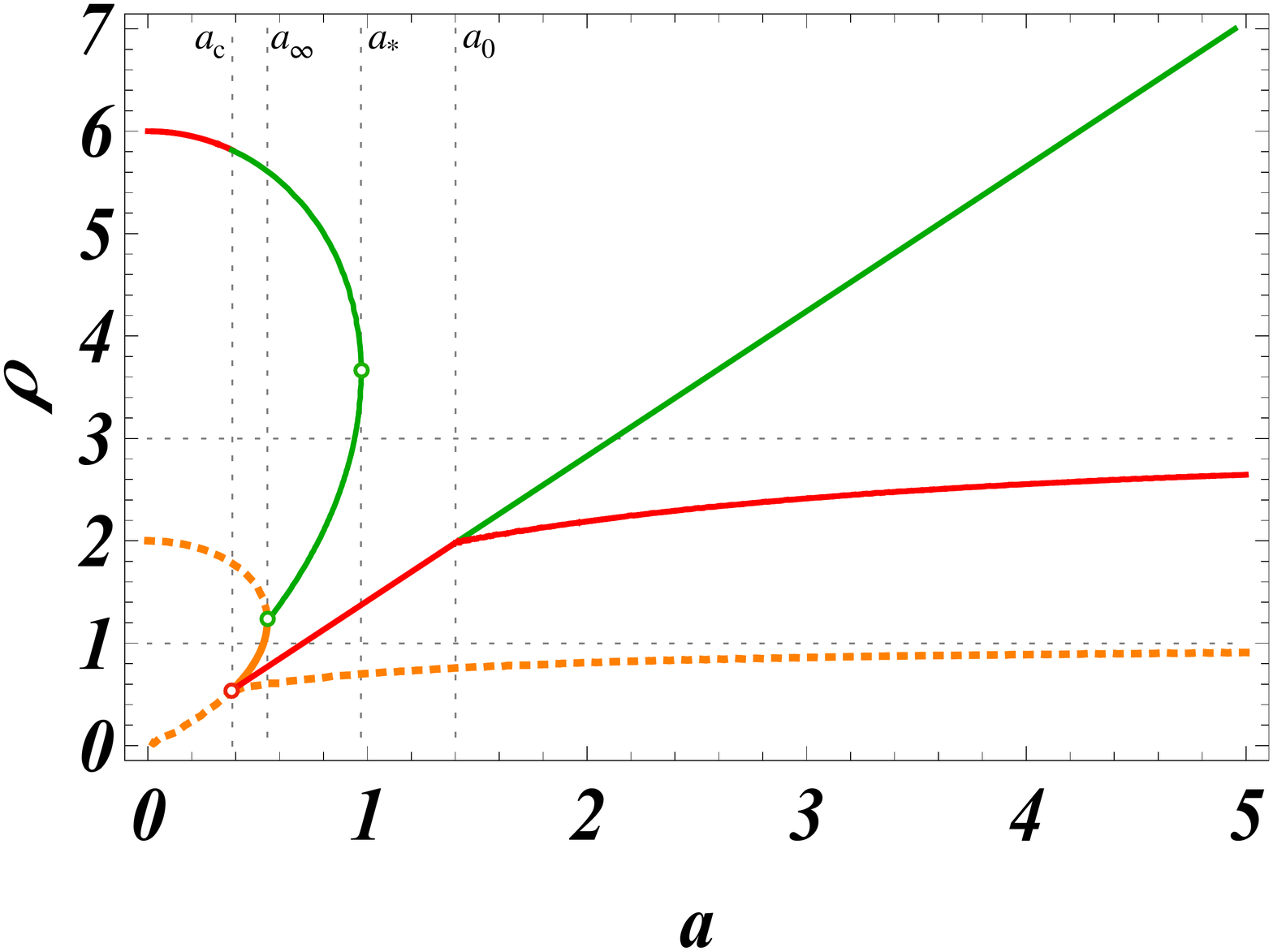}
    \caption{
    Dependence of the radii of marginally stable circular orbits 
    and circular photon orbits on the separation parameter $a$
    in the Majumdar--Papapetrou dihole spacetime with equal unit mass.
    The red and green solid lines mark the radii of the marginally stable circular orbits (MSCOs) and innermost stable circular orbits (ISCOs), respectively.
    The orange dashed lines and the orange 
    solid line are the unstable circular photon orbits 
    and the stable circular photon orbits, respectively. 
     For $a_{\mathrm{c}}<a\leq a_\infty$, 
     the radius of the ISCO is smaller than 
     that of the stable circular photon orbit. 
     At $a=a_\textrm{c}$, a discontinuous transition of ISCO occurs.}
     \label{marginally} 
\end{figure}

\section{Summary and discussions}
\label{IV}
\begin{table}[t]
    \centering
    
    \begin{tabular}{cllclc}
\hline
\hline
&\multicolumn{1}{c}{separation}&\multicolumn{1}{c}{MSCOs}&$n(\mathrm{MSCOs})$&\multicolumn{1}{c}{ISCOs}&$n(\mathrm{ISCOs})$
\\
\hline
\textrm{A}.& $a>a_0=1.401\cdots$ 
&$(\rho_0, z)$ where $h_0(\rho_0, z)=0$, $|z|\leq a$
&$3$&$(\rho_0, z)$ where $h_0(\rho_0, z)=0$, $z\neq0$&$2$\\[1mm]
\textrm{B}.& $a=a_0$
&$(\sqrt{2}a_0,0)$&$1$&$(\sqrt{2}a_0,0)$&$1$\\[1mm]
\textrm{C}.& $a_0>a>a_*=0.9713\cdots$ 
&$(\sqrt{2}a,0)$&$1$&$(\sqrt{2}a,0)$&$1$\\[1mm]
\textrm{D}.& $a=a_*$ 
&$(\sqrt{2}a_*,0)$&$1$&$(\sqrt{2}a_*,0)$&$1$\\[1mm]
\textrm{E}.& $a_*>a>a_\infty=0.5433\cdots$ 
&$(\rho, 0)$ where $h_0(\rho,0)=0$&$3$&$(\sqrt{2}a,0)$&$1$\\[1mm]
\textrm{F}.& $a=a_\infty$ 
&$(\rho, 0)$ where
$h_0(\rho, 0)=0$, 
$0\leq L_0^2 <\infty$
&$2$&$(\sqrt{2}a_\infty,0)$&$1$\\[1mm]
\textrm{G}.& $a_{\infty}>a>a_{\mathrm{c}}=0.3849\cdots$ 
&$(\rho, 0)$ where
$h_0(\rho, 0)=0$, 
$0\leq L_0^2 <\infty$
&$2$&$
(\sqrt{2}a,0)$&$1$\\[1mm]
\textrm{H}.& $a=a_{\mathrm{c}}$ 
&$(\rho, 0)$ where $h_0(\rho, 0)=0$&1&$(\rho, 0)$ where $h_0(\rho, 0)=0$&$1$\\[1mm]
\textrm{I}.& $a_{\mathrm{c}}>a\geq0$ 
&$(\rho, 0)$ where $h_0(\rho, 0)=0$&$1$&$(\rho, 0)$ where $h_0(\rho, 0)=0$&$1$
\\
\hline
\hline
    \end{tabular}
    \caption{Positions in $\rho$-$z$ plane 
    and the numbers of the marginally stable circular orbits~(MSCOs) and the innermost stable circular orbits~(ISCOs) for each range of the separation parameter $a$. The item $n(\mathrm{MSCOs})$ indicates the number of the marginally stable circular orbits, and the item $n(\mathrm{ISCOs})$ 
    indicates the number of the innermost stable circular orbits. }
    \label{magisco1}
\end{table}

We have investigated stable circular orbits 
in the Majumdar--Papapetrou dihole spacetime
with equal unit mass for various values of 
the dihole separation. 
We have divided the separation parameter $a$ into five ranges 
based on qualitative differences of the sequence of
stable circular orbits
and simultaneously have determined the four critical values
as the boundaries of the ranges: 
$a_0, a_*, a_\infty$ and $a_{\textrm{c}}$.
For $a>a_0=1.401\cdots$, 
the sequence of stable circular orbits exists
on the symmetric plane 
and further bifurcates and extends 
towards each black hole. 
This phenomenon is a clear sign to recognize a dihole.
On the other hand, for $0\leq a\leq a_0$, 
stable circular orbits lie only on the symmetric plane. 
In both ranges $a_0\geq a>a_*=0.9713\cdots$ and 
$0\leq a\leq a_{\textrm{c}}=0.3849\cdots$, 
stable circular orbits form a continuous single sequence, 
while in the range 
$a_\textrm{c}<a\leq a_*$,
stable circular orbits form 
two separated sequences.

With the transition of the sequence of stable circular orbits,
the numbers of marginally stable circular orbits changes. 
We have summarized them in Table~\ref{magisco1}. 
The number of the marginally stable circular orbits
increases due to the bifurcation or the separation of the sequence. 
The radii of marginally stable circular orbits and the 
ISCOs 
are plotted as a function of $a$ in Fig.~\ref{marginally}.  
The ISCO radius, shown by red lines, 
can be smaller than the 
ISCO radius
in the single extremal Reissner--Nordstr\"om black hole spacetime. 
The location of the ISCO changes discontinuously at $a=a_{\textrm{c}}$.

For an equal mass MP dihole with arbitrary separation, 
we have found stable circular orbits far from the dihole on the symmetric plane.
These orbits balance by Newtonian gravitational force and centrifugal force. 
Near the dihole, however, stable circular orbits 
may balance by other mechanisms. 
As in the case of a familiar Schwarzschild black hole, 
a particle in the vicinity of the horizon feels 
the high-order relativistic effect. 
On the other hand, since there is no horizon 
on the symmetric plane of this dihole spacetime, 
the centrifugal barrier of a particle inevitably diverges at the center. 
As a result, a radial stable equilibrium point appears by balancing 
the relativistic higher-order gravitational force and centrifugal force. 
Furthermore, if this point is also in a region bounded in the vertical direction, 
a stable circular orbit occurs. 
This mechanics is 
similar to that 
of the appearance of stable circular orbits near 
the 5D black ring~\cite{Igata:2010ye, Igata:2014bga}. 
This suggests that 
the phenomenon occurs universally in the spacetime where there is no
horizon at the center of the system.

We briefly mention 
unstable circular orbits for massive particles.
For an arbitrary value of $a>0$, 
there exists the sequence of 
unstable circular orbits on the symmetric plane 
in the range $\rho<\sqrt{2}a$, which are 
radially stable but 
vertically unstable. 
The sequence further appears 
between the pair of marginally stable circular orbits 
on the symmetric plane for 
$a_\infty<a<a_*$~[see Fig.~\ref{positions}(e)], 
while 
it appears
between the outermost marginally stable circular orbit 
and the unstable circular photon orbit for 
$0\leq a\leq a_\infty$~[see Figs.~\ref{positions}(f)--\ref{positions}(i)]. 
In addition, we also find 
unstable circular orbits on $\rho=\rho_0$ for 
$a>a_{\textrm{c}}$. 
They appear between the ISCO(s) 
and the unstable circular photon orbits. 
On these sequences, 
the energy and the angular momentum are 
monotonically increases as the radius decreases.

It is worthwhile to remark on 
characteristic accretion disk formation 
on the symmetric plane
The two sequences of stable circular orbits for $a_{\textrm{c}}<a<a_*$ suggest 
the formation of 
double accretion disks with a common center. 
In general, 
it is natural for a particle in an accretion disk to 
fall from outside to inside while losing 
their energy and angular momentum. 
If a particle reaches the inner edge of the 
outer disk and loses further 
energy and angular momentum, 
it transits into a stable circular orbit 
in the inner disk, which 
has lower energy and angular momentum 
levels than those of the outer disk. 
Taking into account this mechanism, 
we can obtain a more restricted 
parameter range $a_*>a>0.5238\cdots$ 
from the condition that 
the energy and the angular momentum levels at the 
inner edge of the inner disk 
become smaller than those at the 
inner edge of the outer disk. 
Therefore, we can conclude that 
the formation of double accretion disks occurs in this parameter range.

Though we have considered the MP dihole spacetime 
with equal mass in the present paper, 
the methods 
in the discussions above
are applicable to the 
MP dihole spacetime with different mass. 
Since the MP dihole spacetime is a toy model of a realistic binary system, 
we need to take into account the dynamical effect 
for further understanding of the binary system in future work. 
There are some previous works taking into account the dynamical effect of the binary in the MP dihole spacetime~\cite{Camps:2017gxz, Jai-akson:2017ldo}. 
These works may help us to analyze the dynamical features of the sequence of stable circular orbits in the binary system.

\begin{acknowledgments}
The authors thank
N.~Asaka, T.~Harada, T.~Houri, D.~Ida, H.~Ishihara, T.~Kobayashi, Y.~ Koga, T.~Kokubu, S.~Noda, K.~Ogasawara, H.~Saida, 
and R.~Takahashi for their helpful discussions and useful comments. 
This work was supported by the MEXT-Supported Program 
for the Strategic Research Foundation at Private Universities, 2014--2017 (S1411024),
JSPS KAKENHI Grant No.~JP19K14715~(T.I.) 
and Rikkyo University Special Fund for Research (K.N.). 
\end{acknowledgments}
\appendix
\section{
Circular orbits in the extremal Reissner--Nordstr\"om black hole}
\label{sec:A}
We review 
circular orbits
in the 
extremal Reissner--Nordstr\"om black hole spacetime. 
The metric in the isotropic coordinates is given by
\begin{align}
g_{\mu\nu}\:\!\mathrm{d}x^\mu\:\!\mathrm{d}x^\nu
=-\left(1+\frac{M}{\rho}\right)^{-2} \mathrm{d}t^2+\left(1+\frac{M}{\rho}\right)^2
[\:\!
\mathrm{d}\rho^2
+\rho^2\:\!(\mathrm{d}\theta^2+\sin^2\theta\:\!\mathrm{d}\phi^2)\:\!], 
\end{align}
where $M$ is the mass of the black hole. 
The horizon is located at $\rho=0$. 
The standard form of the metric in the Schwarzschild coordinates 
is recovered by the transformation $r=\rho+M$. 
We consider a freely falling 
particle in a circular orbit on the equatorial plane ($\theta = \pi /2$). 
Using the same procedure as in Sec.~\ref{II}, 
we obtain the radial equation
\begin{align}
&\dot{\rho}^2+V=E^2,
\\
&V=\frac{L^2}{\rho^2}\left(1+\frac{M}{\rho}\right)^{-4}+\kappa\left(1+\frac{M}{\rho}\right)^{-2},
\end{align}
where the dot denotes derivative with respect to an affine parameter, 
$E=-(1+M/\rho)^{-2}\:\!\dot{t}$ is conserved energy,
$L=(\rho+M)^2\dot{\phi}$ is conserved angular momentum, 
$\kappa=1$ for massive particles, and $\kappa=0$ for massless particles.
Circular orbits are given by 
the simultaneous solutions of $V=E^2$ and $\mathrm{d}V/\mathrm{d}\rho=0$. 
The second equation for massless particles (i.e., $\kappa=0$) has a root 
\begin{align}
\rho=M. 
\end{align}
This is the radius of the circular photon orbit. 
In the Schwarzschild radial coordinate, the radius corresponds to $r=2M$.
There is the other root $\rho=0$ but this is not a circular orbit. 
On the other hand, the equation $\mathrm{d}V/\mathrm{d}\rho=0$ for massive particles (i.e., $\kappa=1$)
has the roots
\begin{align}
\rho_\pm=\frac{L^2-2M^2\pm\sqrt{L^2(L^2-8M^2)}}{2M}.
\end{align}
Note that the larger root $\rho_+$ is a local minimum point of $V$, 
while the smaller root $\rho_-$ is a local maximum point of $V$. 
Thus, stable circular orbits exist at $\rho=\rho_+$, 
and unstable circular orbits exist at $\rho=\rho_-$. 
In the case $L^2=8M^2$, the pair of circular orbit radii coincides with each other, and then 
the radius takes the value
\begin{align}
\rho=3M. 
\end{align}
This is the minimum radius of stable circular orbits and is known as 
the innermost stable circular orbit. 
In the Schwarzschild radial coordinate, 
the radius of the innermost stable circular orbit corresponds to $r=4M$.

\section{Circular photon orbits
in the equal mass Majumdar--Papapetrou dihole spacetime}
\label{sec:B}
We review 
circular photon orbits
in the MP dihole spacetime with equal unit mass $M_\pm=1$. 
The effective potential for null particles
is given by Eq.~\eqref{potential} with $\kappa=0$. 
As is the case with timelike particles, 
the condition $V_z=0$ is equivalent to $U_z=0$ and 
has solutions $\rho = \rho_0$ and $z=0$.
We focus on 
circular photon orbits
on $\rho=\rho_0$. 
To investigate 
their
positions, 
we solve Eq.~\eqref{modiUz} again for $z$ and
find real 
roots $z_0(\rho)$.
The line $z=z_0$ corresponds to 
the line $\rho=\rho_0$.
We consider the condition $V_\rho =0 $ on 
$z=z_{0}$:
\begin{align}
V_\rho(\rho, z_{0}(\rho))=0.
\end{align}
The real solutions of this equation express the radii
of the unstable circular photon orbits. 
They only exist for  $a\geq a_{\textrm{c}}$ 
but not for  $a<a_{\textrm{c}}$.
The dependence of the radii of circular photon orbits 
on the separation parameter $a$ is 
shown in Fig.~\ref{marginally}
by orange dashed lines.
In the limit as $a \to \infty$, the circular photon orbit radius measured by $\rho$ approaches 1, 
which coincides with that
of the single Reissner--Nordstr\"om black hole spacetime.

Next we focus on circular photon orbits on
$z=0$ plane.  The condition $V_ \rho =0$ leads to 
  \begin{align}
\label{eq:nullVrho}
    \left( \rho ^ { 2 } + a^2 \right) ^ { 3 / 2 } 
    = 2 \left( \rho ^ { 2 } - a^2 \right).
  \end{align}
This equation has real roots only for $\rho > a$.
We can rewrite Eq.~\eqref{eq:nullVrho} 
\begin{align}
f=0,
\label{nullVr}
\end{align}
where $f$ is defined by Eq.~\eqref{eq:f}.
In the range $\rho>a$, 
the roots of this cubic equation for 
$\rho^2$ are
given by
  \begin{align}
  \label{eq:rhopu}
    \rho^2_{\mathrm{pu}}
    &= 
     \frac{4 - 3 a^2}{3}+\frac{8}{3} \sqrt{1-3 a^2} \cos \left[\:\!
    \frac{1}{3} \arccos 
    \frac{27 a^4-36 a^2+8}{8 \left(1-3a^2\right)^{3/2}}\:\!
    \right], \\
\label{eq:rhops}
\rho ^2_{\mathrm{ps}}
&=\frac{4 - 3 a^2}{3}+\frac{8}{3} \sqrt{1-3 a^2} \cos \left[\:\!
\frac{4\pi}{3} + \frac{1}{3} \arccos 
\frac{27 a^4-36 a^2+8}{8 \left(1-3a^2\right)^{3/2}}
\:\!\right]. 
  \end{align}
Note that 
$\rho _{\mathrm{pu}}$
and 
$\rho _{\mathrm{ps}}$
correspond to the 
unstable circular photon orbit
and the stable one, respectively~(see Fig.~\ref{marginally}).
These roots are real 
only for $a \leq a_\infty$, 
which is 
found from the discriminant of Eq.~\eqref{nullVr}. 
Hence, 
for $a>a_\infty$,
there exist unstable 
circular photon orbits 
only on
$\rho=\rho_0$ line~[see Figs.~\ref{positions}(a)--\ref{positions}(e)]. 
When $a=a_\infty$, 
an additional circular photon orbit appears 
at $(\rho, z)=(\rho_\infty, 0)$~[see Fig.~\ref{positions}(f)], 
where $\rho_\infty$ is given by Eq.~\eqref{eq:rhoinf}.
The condition for 
the existence of the multiple root
$\rho_{\mathrm{pu}}=\rho_{\mathrm{ps}}$ 
also leads to the values of $a_\infty$ and $\rho_\infty$.
For $a_{\textrm{c}}<a<a_\infty$, 
unstable circular photon orbits 
exist on $\rho=\rho_0$ and at 
$(\rho, z)=(\rho_{\textrm{pu}},0)$, and 
stable circular photon orbits exists 
at $(\rho, z)=(\rho_{\textrm{ps}},0)$~[see Fig.~\ref{positions}(g)].
When $a=a_\mathrm{c}$, three circular photon orbits degenerate 
at $(\rho,z)=(\sqrt{2}a_{\textrm{c}},0)$~[see Fig.~\ref{positions}(h)].
For $a<a_{\textrm{c}}$, there is no stable circular photon orbit 
but there are two unstable circular photon orbits on $z=0$.
The outer one is radially unstable but vertically stable, while 
the inner one is radially stable 
but vertically unstable~(see also Ref.~\cite{Shipley:2016omi}). 
For $a=0$, we obtain $\rho_{\textrm{pu}}=2$ 
and $\rho_{\textrm{ps}}=0$, which coincide with 
the radius of the unstable circular photon orbit 
and the horizon radius of the
single extremal Reissner--Nordst\"om black hole with mass 2, 
respectively~(see Appendix~\ref{sec:A}).

\end{document}